\documentclass[12pt,a4paper,final]{iopart}
\usepackage[utf8]{inputenc}
\usepackage{graphicx,color}% Include figure files
\usepackage{dcolumn}% Align table columns on decimal point
\usepackage{bm}% bold math
\usepackage[breaklinks=true,colorlinks=true,linkcolor=blue,urlcolor=blue,citecolor=blue]{hyperref}
\usepackage{iopams}  
\begin{document}
\title[Effect of intrinsic defects on the thermal conductivity of PbTe]{Effect of intrinsic defects on the thermal conductivity of PbTe from classical molecular dynamics simulations}
\author{Javier F. Troncoso, Pablo Aguado-Puente and Jorge Kohanoff}
\address{Atomistic Simulation Centre, Queen's University Belfast, Belfast BT7 1NN, UK}%
\eads{\mailto{jfernandeztroncoso01@qub.ac.uk}, \mailto{p.aguadopuente@qub.ac.uk}, \mailto{j.kohanoff@qub.ac.uk}}
\begin{abstract}
Despite being the archetypal thermoelectric material, still today some of the most exciting advances in the efficiency of these materials are being achieved by tuning the properties of PbTe. Its inherently low lattice thermal conductivity can be lowered to its fundamental limit by designing a structure capable of scattering phonons over a wide range of length scales. Intrinsic defects, such as vacancies or grain boundaries, can and do play the role of these scattering sites. Here we assess the effect of these defects by means of molecular dynamics simulations. For this we purposely parametrize a Buckingham potential that provides an excellent description of the thermal conductivity of this material over a wide temperature range. Our results show that intrinsic point defects and grain boundaries can reduce the lattice conductivity of PbTe down to a quarter of its bulk value. By studying the size dependence we also show that typical defect concentrations and grain sizes realized in experiments normally correspond to the bulk lattice conductivity of pristine PbTe. 
\end{abstract}
\submitto{\JPCM}

\maketitle

%\tableofcontents

\section{Introduction}\label{sec:level1} 
%\pa{We need to cut the intro}

The utilization of a temperature gradient to create an electrical current through the Seebeck effect has long been seen as a potentially revolutionary way of producing energy. 
Nowadays, efforts to improve the performance of thermoelectric materials are mostly motivated by the possibility of using them to convert waste heat into usable energy \cite{Schierning2018}.
Unfortunately the efficiency of currently available thermoelectric materials is typically too low for any but a handful of real-life niche applications as recently reviewed by Champier \cite{Champier2017}.
They are ideal as compact and robust solid state generators for applications requiring very high reliability, such as space satellites and probes \cite{Schmidt2009}.
%
% The effect can be used to design solid state generators, compact and robust, ideal for applications requiring very high reliability, such as space satellites and probes \cite{Schmidt2009}.

The thermodynamic efficiency of a thermoelectric is measured by its dimensionless figure of merit, $ZT$, defined as
\begin{eqnarray}
ZT=\frac{S^2 \sigma T}{\kappa},
\label{eq:ztdimensionless}
\end{eqnarray}
where $S$ is the Seebeck coefficient, $\sigma$ is the electrical conductivity, $T$ is the absolute temperature and $\kappa$ is the thermal conductivity due to both electrons and phonons. For any practical application it is typically considered that $ZT$ must be close to or larger than 1, with reports of $ZT>2$ considered as breakthroughs in the field.

One of the most studied and used materials for thermoelectric applications is lead telluride, PbTe. This small gap semiconductor possesses a remarkable figure of merit at intermediate temperatures even in its pristine form ($\sim 0.8$ at 700 K)\cite{Snyder2008}. The high performance of PbTe is largely due to its low lattice thermal conductivity, with $\kappa_\mathrm{lat}\sim 2$ Wm$^{-1}$K$^{-1}$ at 300 K  \cite{Snyder2008}, which in turn stems from peculiarities in the phonon dispersion and scattering properties of the material.
Inelastic neutron scattering experiments \cite{Delaire2011} as well as first principles calculations \cite{An2008a, Shiga2012,Ju2018} have shown that the lattice dynamics of this material is characterized by a strong anharmonic coupling between the longitudinal acoustic (LA) and transverse optical (TO) branches. This coupling leads to an anomalously large dependence of the TO mode frequency on the volume \cite{An2008a} and, more importantly, to a strong damping and lowering of the LA branch \cite{Delaire2011,Shiga2012}. This, together with the relatively low group velocity of transverse acoustic (TA) modes \cite{Cochran1966}, are two of the main reasons behind the characteristically low values of $\kappa_\mathrm{lat}$ of PbTe.  

% According to inelastic neutron scattering experiments \cite{Delaire2011}, X-ray scattering \cite{Jiang2016} and calculations \cite{Shiga2012, Shiga2014}, acoustic branches, particularly the transversal acoustic (TA), are extremely anharmonic, and the group velocity of the long wave TA modes is very low. The lattice thermal conductivity depends directly on this group velocity, so such slow propagation speed of active carriers is expected to limit the contribution of TA phonons to the overall thermal conductivity \cite{Shiga2012}. On the other hand, the mode Grüneisen parameter indicates that the dynamics of transverse optical (TO) phonons near $\Gamma$ point is also very anharmonic, and this is expected to enhance scattering rates with acoustic phonons \cite{Shiga2012}.  The frequency of this soft $\Gamma$ mode increases with increasing temperature \cite{Delaire2011} and the energy of these TO phonons near the zone centre ($\Gamma$ point) is highly sensitive to changes in volume and pressure \cite{Shiga2012}.  Long-wavelength acoustic modes are essentially pressure waves, so this dependence of TO modes on the pressure suggests a strong coupling between TO and acoustic phonons, and inelastic neutron scattering experiments show that this coupling is with longitudinal acoustic (LA) phonons \cite{Delaire2011}. 

Efforts to increase the intrinsic thermoelectric performance of PbTe-based materials have led to record-high values of the figure of merit. Strategies oriented towards tuning the electronic structure of the system have produced samples with $ZT$ values in excess of 2. Such strategies include populating the high-multiplicity $\Sigma$ hole pockets via Na-doping \cite{Pei2011b}, introducing resonant defect levels by Tl-doping\cite{Heremans2008}, or using a combination of alloying with SeTe and Na doping in order to align the $\Gamma$ and $\Sigma$ hole-pockets \cite{Pei2011}. 
An alternative approach is to tackle the lattice contribution to the thermal conductivity. The preferred approach in this case has typically been the use of nanostructuring \cite{Pei2011c,Biswas2012,Zhao2014b} as a means to scatter phonons over a wide range of length scales. Recently the possibility of driving the material closer to the ferroelectric phase transition by alloying with GeTe has been proposed as a method to enhance the LA-TO scattering rates \cite{Murphy2016,Murphy2017}. 
%\pa{is this alloy really feasible? see eutectic system by Sootsman 2009}

Despite the great attention devoted to PbTe, the role of \emph{intrinsic} defects on its thermal transport properties has been scarcely investigated. Experimentally, this is partly due to the difficulty in controlling and estimating the concentrations of neutral and stoichiometric point defects which, according to calculations \cite{Li2015} and experiments \cite{Schenk1988}, may actually be the most common in the pure material. First-principles simulation studies, on the other hand, have addressed the energetics \cite{Bajaj2015,Goyal2017} and electronic properties \cite{Li2015,Goyal2017} of point defects, but their influence on the lattice dynamics as well as the role of other types of intrinsic defects such as grain boundaries are still difficult to tackle within this framework.
The difficulty in minimizing the lattice contribution to the thermal conductivity lies in the very wide range of time and length scales of the processes involved. 
%
% MD simulations allows to reproduce scattering processes in bulk, but sb initio modeling is computationally demanding and the calculations of lattice thermal conductivities are not straightforward. On the other hand, classical molecular dynamics calculations are advantageous in modeling heat transfer across various geometries in short runs.
%
As such, one of the essential theoretical tools to investigate these materials is molecular dynamics (MD) simulation. Classical MD simulations have been used before to explore thermal transport in PbTe and the effect of different strategies aimed at reducing the lattice conductivity. 
In \cite{Chonan2006}, an effective Coulomb potential fitted to reproduce experimental lattice parameters was employed to study the effect of alloying PbTe with the ferroelectric materials SnTe and GeTe, showing a remarkable decrease in lattice thermal conductivity. This result is in agreement with later density functional theory (DFT) calculations which showed that the decrease was the result of the softening of the TO modes of the material and the consequent enhanced scattering of acoustic phonons \cite{Murphy2016}. 
A Coulomb-Buckingham potential developed by Qiu {\it et al.}, which was fitted to reproduce the Murnhagan equation of state of bulk PbTe obtained with DFT, shows a significant reduction of $\kappa_\mathrm{lat}$ when relatively large concentrations of vacancies ($>2\cdot 10^{-3}$) are introduced in the system \cite{Qiu2012}. The same potential was also utilized to investigate the effect of twin boundaries on the thermal transport in PbTe \cite{Zhou2018}. 
In \cite{Kim2012a}, a combination of classical and {\it ab initio} MD was used to determine the effect of the strong anharmonicity of vibrations in PbTe on the electronic structure of the material and therefore on its thermoelectric figure of merit.  
The anomalous lattice dynamics observed in neutron diffraction experiments \cite{Bozin2010} was also interpreted as the result of a very large anharmonicity in short-range interactions \cite{Shiga2012,Shiga2014}. This investigation was carried out using MD simulations with interatomic force constants (IFC) computed with DFT. The same methodology was used to investigate thermal conductivity in PbTe, PbSe and PbTe$_{1-x}$Se$_x$ \cite{Shiga2012,Tian2012}. However, even {\it ab initio} methods have difficulties, arising from the non-trivial temperature dependence of the potential energy surface, to quantitatively predict the thermal properties of these materials from zero-temperature IFC, as pointed out by Romero {\it et al.}\cite{Romero2015} using a more sophisticated method, consisting in the sampling of the temperature-dependent IFC through fully {\it ab initio} MD simulations.  

Despite their important contributions to the understanding of lattice thermal conductivity in PbTe, all currently available classical potentials \cite{Chonan2006,Qiu2008,Kim2012a} tend to overestimate the lattice thermal conductivity of the bulk material (see section \ref{sec:methods}) and therefore lack quantitative predictive power. This overestimation has been attributed to the extra scattering provided by defects and grains boundaries in experimental samples \cite{Qiu2012}. However, the calculation of the contribution arising from these mechanisms for realistic defect concentrations has never been tackled and, therefore, remains at the level of speculation.

Available force fields have been fitted to reproduce the structure and some mechanical properties of the material. Instead, we develop a Coulomb-Buckingham potential that reproduces also features of the phonon dispersion of PbTe, hence improving the values of thermal conductivity. We then use this potential to study, via MD simulations, the influence of defects on the thermal conductivity of PbTe under realistic conditions. We consider vacancies, interstitial atoms and grain boundaries.

\section{Molecular dynamics simulations of bulk lead telluride}\label{sec:methods} 

Various classical interatomic potentials have been proposed for this material at intermediate temperatures \cite{Qiu2012,Kim2012a,Chonan2006}. They were all obtained by fitting structural properties and, when the size of the simulation box is converged, they all tend to overestimate the lattice thermal conductivity by $\sim 50\%$. More sophisticated methods such as {\it ab initio} MD or dynamics based on {\it ab initio} interatomic force constants usually perform better, but at a much larger computational cost that prevents the simulation of materials with realistic defects. 

Here we propose a new parametrization of a Coulomb-Buckingham potential fitted to reproduce a selection of phonon frequencies, structural, and mechanical properties of PbTe. This choice is dictated by our aim of improving on thermal transport properties over currently available models. The expression of the Coulomb-Buckingham potential is
\begin{eqnarray}
V \left( r_{ij} \right) = \frac{q_i q_j}{r_{ij}} + A_{ij} e^{-r_{ij}/\rho_{ij}}-\frac{C_{ij}}{r_{ij}^6},
\label{eq:Buckingham}
\end{eqnarray}
where the parameters $q_i$ are the effective ionic charges, $A_{ij}$ and $\rho_{ij}$ control the hardness of the repulsive barrier and $C_{ij}$ modifies the short-range attractive part. Note that this potential, best suited for ionic systems, is repulsive between equal chemical species, whereas the interaction is attractive between different atom types, as expected. This is largely thanks to the presence of the Coulomb term, in which the charges were included as fitting parameters in the model. The fitting procedure was performed using the Broyden-Fletcher-Goldfarb-Shanno (BFGS) method in the GULP code \cite{Gale1997,Gale2003}. 
%In order to capture thermal transport properties we chose to fit our potential to reproduce a selection of phonon frequencies and structural parameters. 

We used the optical phonon frequencies in order to capture the anomalous behavior of TO modes and to restrict the frequency range of the phonon dispersion. Since this type of potential does not account for the polarizability of the electronic cloud, the LO-TO splitting of materials with large dynamical charges, such as PbTe, is not expected to be accurately reproduced \cite{Zhang2011a,Gonze1997}. Therefore we chose the frequencies at the X-point in the Brillouin zone (zone boundary) as a reference, instead of those at the $\Gamma$-point.
 For relatively simple interatomic potentials, the choice of fitting phonon frequencies has the undesired effect of a loss of accuracy in the elastic properties, and vice versa \cite{Powell2006}. Therefore, both phonon frequencies and elastic properties cannot be fitted simultaneously and a compromise needs to be made. While fitting to elastic properties is suitable if the force field is to be used solely for structural relaxation, the use of phonon frequencies is more suitable for the study of thermal properties due to their link to the acoustic modes and sound velocities.
Structural and mechanical properties were taken from experiment, while the phonon frequencies were obtained from density functional perturbation calculations as implemented in the Quantum Espresso package \cite{Giannozzi2009} and experiment. DFT simulations were performed using PBEsol, a $60$ Ry cutoff for the basis of plane waves, a $12\times 12\times 12$ Monkhorst-Pack sampling of the Brillouin zone \cite{Monkhorst1976} and norm-conserving pseudpotentials. Spin-orbit coupling was switched on. %this was a footnote

\begin{figure}[h!]
\centering
\includegraphics[width=\columnwidth]{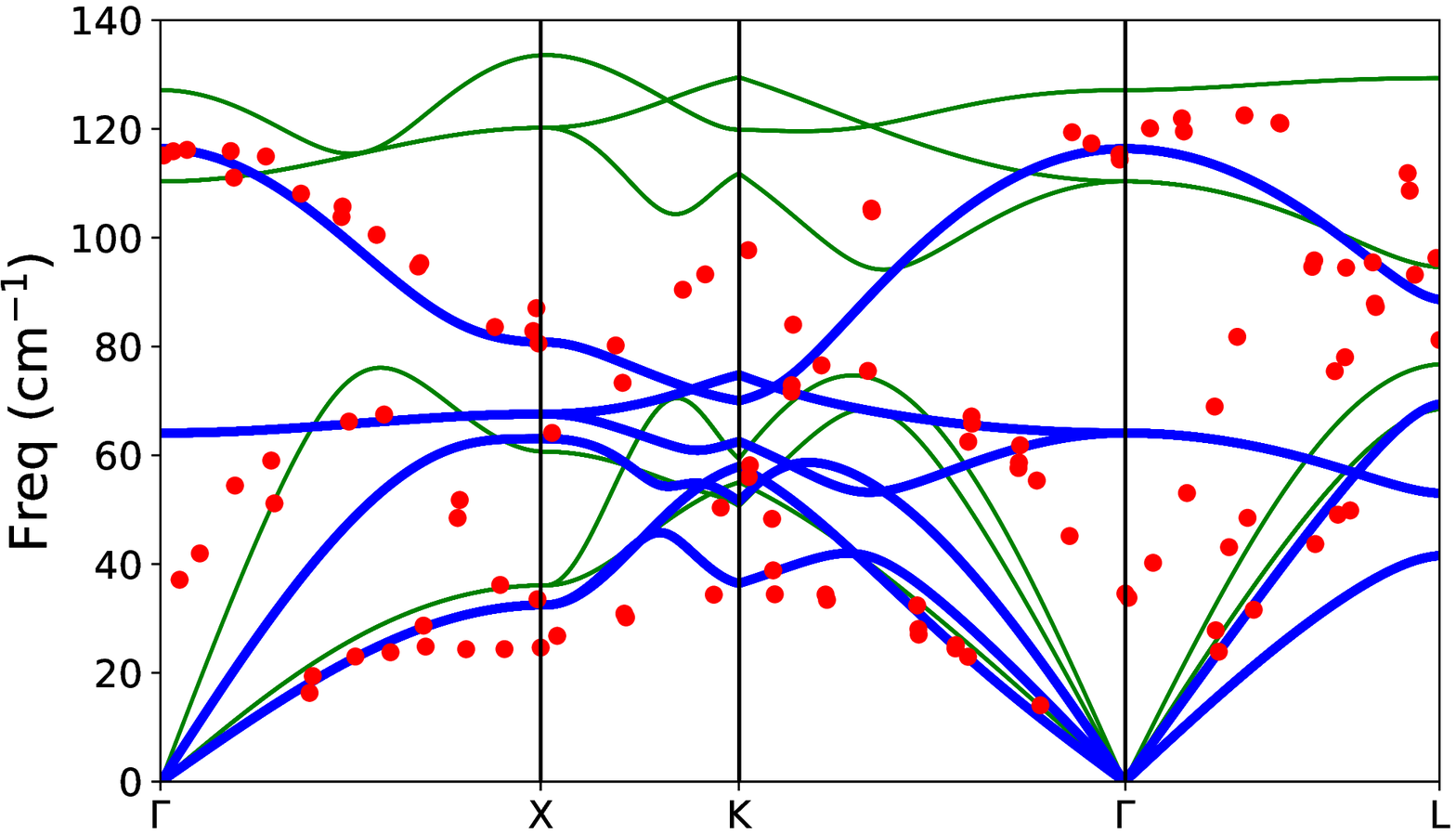}
\caption{ Phonon dispersion relations obtained with the present model (solid blue lines), compared to the model proposed in \cite{Qiu2012} (solid, thin green lines) and experimental values \cite{Cochran1966} (red dots).}
\label{fig:dispersion}
\end{figure}

The parameters of the resulting potential are listed in table  ~\ref{table:mybuckparams}. The potential is benchmarked against a selection of fitted and non-fitted phonon frequencies and structural and mechanical properties in table \ref{table:fittingparams}. As expected, a good agreement was obtained for the fitted physical properties. 
More importantly, the restriction of the frequency range of the phonon dispersion at X leads to a remarkable improvement in the acoustic modes and the Debye temperature. Phonon  dispersion relations were compared with Qiu’s \cite{Qiu2012} and experimental dispersions \cite{Cochran1966} and show a noticeable improvement (see figure \ref{fig:dispersion}). 

%{\color{red} The phonon frequencies were fitted  accurately  but  at  the  expenses  of  the  elastic properties. D. Powell et al. \cite{Powell2006} showed that it is not possible to simultaneously fit phonon frequencies and elastic constants, and therefore a compromise has to be made. While the use of the elastic constants is necessary if  the  main  goal  is  to  use  the  new  potential  solely  for  structural  relaxation, the fitting to phonon frequencies is mandatory if the objective is to study phonon properties. Due to the close link of the thermal conductivity with full phonon spectrum, especially acoustic branches phonons, and the sound velocities in the material, we decided on the use of phonon frequencies as fitting parameters despite the loss of accuracy in the elastic constants.} 

\begin{table}[h!]
\caption{\label{table:mybuckparams} Potential parameters. Partial charges are  $q_\mathrm{Pb}$=+1.004848 and  $q_\mathrm{Te}$=-1.004848.} 
\begin{indented}
\item[]
\lineup
\begin{tabular}{@{}lccc}
\br
& $A$ (kcal$\cdot$mol$^{-1}$) & $\rho$ (\AA) & \textit{C} (kcal$\cdot$\AA$^6 \cdot$mol$^{-1}$) \\
\mr
Pb-Pb & 715 & 0.279 & 8.51$\cdot 10^{-3}$ \\
Pb-Te & 86695 & 0.313 & 3.83$\cdot 10^{-3}$ \\
Te-Te & 2966 & 9.25$\cdot 10^{-2}$ & 36.3 \\
\br
\end{tabular}
\end{indented}
\end{table}

 \begin{table}
\caption{\label{table:fittingparams}Selection of physical magnitudes used as benchmark as well as fitting parameters for generating the classical potential. All reference values were obtained from experiments \cite{MadelungHandbook_PbTe,Miller1981}, except for the phonon frequencies that have been obtained here from calculations using density functional perturbation theory (see text for details). The parameters marked with an asterisk are the ones used to perform the fitting with GULP \cite{Gale1997,Gale2003}. The thermal expansion  reported value is for the linear thermal expansion at 300 K. More details regarding the sound velocity in the Supplemental Material. The vacancy formation energy is compared with first-principle calculations \cite{WunFanLi2015}. }
\begin{indented}
\item[]
\lineup
\begin{tabular}{@{}lcc}
\br
 & Ref. values & Potential \\
\mr
Lattice constant, $a_0$  (\AA) & 4.569* & 4.552 \\
Bulk modulus (GPa)      & 39.8 & 17.5  \\
Shear modulus (GPa)    & 21.4 & 10.5  \\
Thermal expansion  ($10^{-6}$ K$^{-1}$) & 20 & 41\\ 
\mr
Phonons at $\Gamma$ (cm$^{-1}$) & & \\
 TO  & \032 & \064 \\%\19
 LO  & 114 & 116 \\%\19
Phonons at X (cm$^{-1}$) & & \\
TA & \024 & \030 \\%\27
LA & \030 & \061 \\ %\39
TO & \0\069* & \068 \\%\76
LO & \0\081* & \081 \\%\92
\mr
Elastic constants (GPa)& & \\
$C_{11}$ & 105 & 37 \\
$C_{12}$ & \0\0\0\07.0* & \0\07.6 \\
$C_{44}$ & \0\0\013.2* & \0\07.6\\
\mr
Debye temperature $\theta_\mathrm{D}$ (K) & 177 & 168\\
\mr
Sound velocities (km/s)\\
$v_{LA}$&2.98 & 2.6\\
$v_{TA}$&1.74 & 1.7\\
\mr
Vacancy formation energy (eV)& & \\
Schottky dimer&1.21 & 1.40\\
Schottky pair&1.68 & 2.40\\
\br
\end{tabular}
\end{indented}
\end{table}

Lattice thermal conductivities of bulk PbTe were calculated using the Green-Kubo method applied to equilibrium MD simulations \cite{Zwanzig1964,Che2000}, except for the simulations of single grain boundaries (see section \ref{sec:grains}). MD simulations were carried out using the LAMMPS code \cite{Plimpton1995,lammpsWeb} for a $12\times 12\times 12$ simulation box, using an integration time step of $2$ fs at 500 K, which was suitable to converge time correlation functions (different time steps were used in the simulations but maintaining the same correlation time of 100 ps). The structure was first simulated in the NVT ensemble for $750$ ps to ensure equilibration, and then the heat current was extracted from the following $9$ ns. Correlation time and simulation box size were increased for simulation cells including defects in order to ensure convergence. A typical heat current autocorrelation function is shown in figure S4 of the supplementary material, together with details of the MD simulations.
%\jf{Simulations details have been included: timestep=3fs, 250m+3M timesteps.}

\begin{figure}[h!]
\centering
\includegraphics[width=\columnwidth]{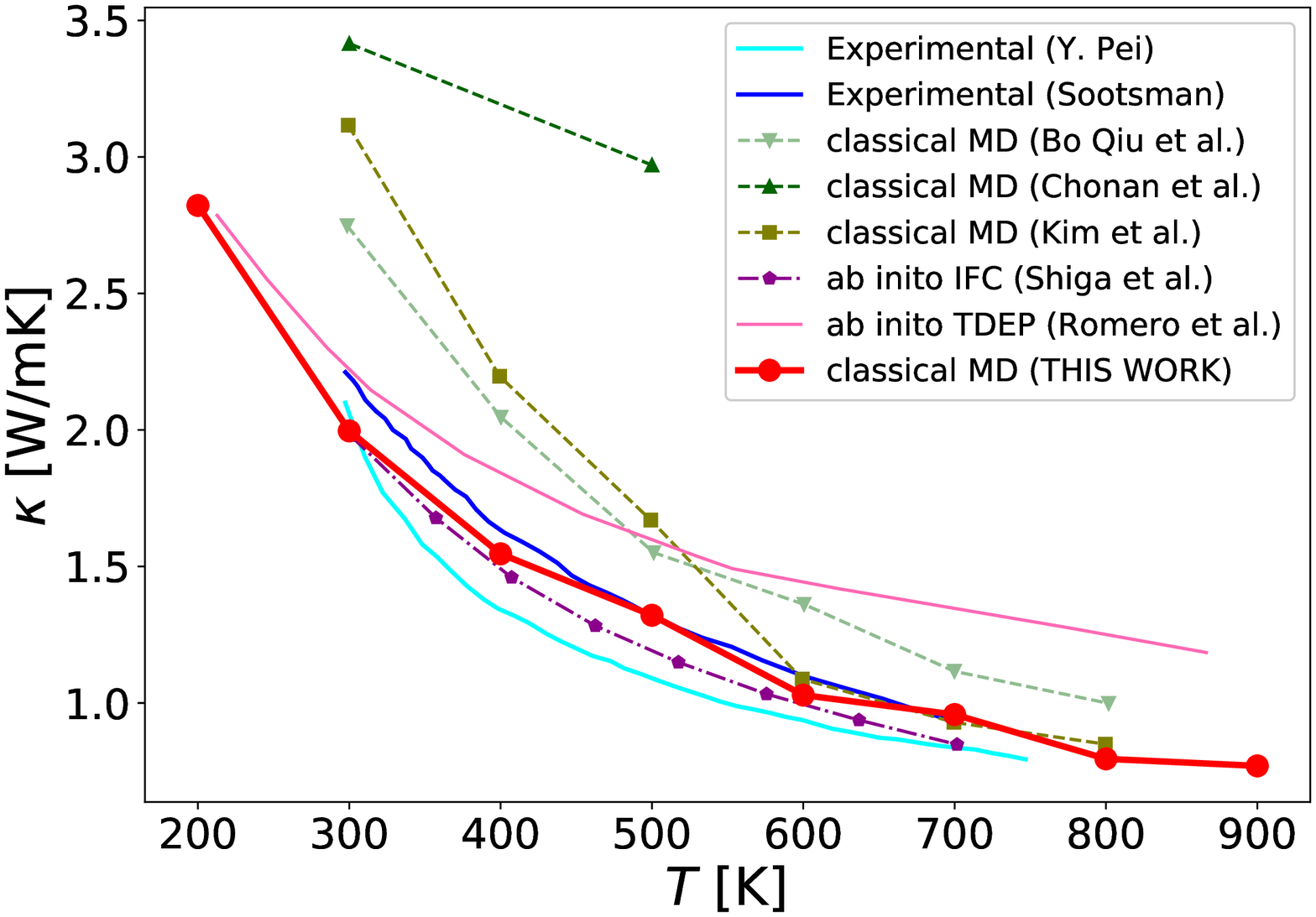}
\caption{Lattice thermal conductivity in the temperature interval 300-800 K obtained with the classical potential presented in this work (thick red line). Results of this work are compared with a selection of experimental and theoretical results.} 
\label{fig:myconductivity}
\end{figure}

%In this method, equilibrium molecular dynamic simulations are performed in order to calculate the local heat fluxes due to instantaneous fluctuations in temperature. The Green-Kubo method relates  the  lattice  thermal  conductivity  of  the  system  to  the  time  required  for such fluctuations to dissipate. Then, in this method, the thermal conductivity is computed by integrating the heat current autocorrelation function using the Kubo linear-response formula \cite{phillpot}:
% \begin{eqnarray}
% \kappa_{\mu \nu}=\frac{1}{k_B T^2 V} \int _0 ^\infty dt' < J_\mu (t'=0)J_\nu (t') >
% \end{eqnarray}
% where $V$ is the volume of the simulation cell, $J$ is the instantaneous heat current and $T$ is the absolute temperature.  \par
% Usually, it is difficult to calculate this instantaneous heat current, but for pair potentials we can use the following expression:
% \begin{eqnarray}
% \vec{J}\equiv \frac{d}{dt} \sum _i \vec{r_i}E_i=\sum _i \vec{v_i}E_i+\sum _i \vec{r_i}\frac{d}{dt} E_i \equiv \vec{J_{kin}}+\vec{J_{pot}}
% \end{eqnarray}
% where $m_i$ , $\vec{v_i}$ , and $E_i$ are the mass, velocity and total energy of particle $i$, respectively.
%
As shown in figure~\ref{fig:myconductivity} we obtained thermal conductivities in excellent agreement with experiments for the temperature interval 300-800K, which is the range of interest for this material. Our parametrization clearly improves over the previously available classical potentials, producing a curve as good as the one obtained with more sophisticated methods, such as the MD based on {\it ab initio} IFC by Shiga {\it et al.} \cite{Shiga2012,Shiga2014}. 
It is important to remark that the two sets of experimental thermal conductivities reported in figure~\ref{fig:myconductivity} (blue and cyan lines) \cite{Pei2011,Sootsman2009} agree within 10-20\%. The values obtained using the present model (red line) fall comfortably within this range.

% These calculations were obtained with MD simulations at constant NVT run for 25 ns after equilibration to keep the temperature constant, but no changes were observed in runs at constant NVE. Red dots stand for the lattice thermal conductivity at a fixed lattice parameter, and are compared with experimental values.

Another benchmark can be made against the Gr\"uneisen parameter, $\gamma$. At high temperature it can be assumed than anharmonic phonon-phonon interactions dominate phonon scattering, and hence the lattice thermal conductivity can be expressed as \cite{Slack1979,Morelli2002}  
\begin{eqnarray}
\kappa_\mathrm{lat} =   A_\mathrm{\kappa} \frac{\bar{M} \theta_\mathrm{D}^3 \delta}{\gamma^2 n^{2/3} T},
\label{eq:gruneisen}
\end{eqnarray}
where $\bar{M}$ is the average atomic mass, $\theta_\mathrm{D}$ is the Debye temperature, $n$ is the number of atoms in the primitive cell and  $\delta$ is equal to $(a_0^3/4)^{1/3}$, where $a_0$ is the lattice constant of the fcc unit cell. When $\bar{M}$ is expressed in atomic mass units, $\delta$ in {\AA} and the thermal conductivity in W m$^{-1} K^{-1}$, the constant $A_\mathrm{\kappa}$ \cite{Slack1979} takes the value $\approx 3.04\cdot 10^{-6}$. Fitting the red curve in figure \ref{fig:myconductivity} with the expression (\ref{eq:gruneisen}) we obtain a Gr\"uneisen parameter $\gamma\approx 2.1$, close to the experimental value of 2 \cite{Zhang2009}, thus confirming the good performance of this new potential in reproducing the lattice thermal properties of bulk PbTe over a wide range of temperatures. Further benchmarks that verify the suitability of the model, including the quality of the forces, comparison of energies between different strutures and dispersion relations for the Gr\"uneisen parameters, are presented in Sec. IA and IB of the supplementary material.

% $E$ the energy and $V$ is the volume. Note that this constant can be given as a function of the  thermal expansion coefficient, $\alpha$, the isothermal bulk moduli, $k_T$, the heat capacity, $C_V$, and the density $\rho$. Above the Debye temperature, the thermal fluctuations of density can be analyzed in terms of phonon-phonon iterations, and it is possible to express this constant as a function of the lattice thermal conductivity \cite{ziman,leethemoelectrics}:

% \begin{eqnarray}
% \kappa_l =   \frac{k_B}{2\pi^2v_s}\left(\frac{k_BT}{\hbar}\right)^3\int_0^{\theta_\mathrm{D}/T} \tau \frac{x^4e^x}{(e^x-1)^2}dx
% \end{eqnarray}
% where $v$ is the velocity of sound, $x=\hbar\omega=k_BT$ and $\tau$ is the Umklapp relaxation time \cite{leethemoelectrics}:

% \begin{eqnarray}
% \tau^{-1}=\frac{380\pi}{81}N_A\hbar\left(\frac{6\pi^2}{4}\right)^{1/3}\frac{\gamma^2}{M_{AB}\delta^2}\left(\frac{T}{\theta_\mathrm{D}}\right)^3x^2
% \end{eqnarray}
% where $N_A$ is the Avogadro’s number, $M_{AB}$ is the average atomic mass of the compounds and $\delta$ is the mean atomic size.

% With this expression, we get to a value of $2.4$, which is slightly higher than the experimental value ($\approx 2$).

\section{Strategies to reduce the lattice thermal conductivity}\label{sec:results} 

Since the seminal works of Dresselhaus and coworkers \cite{Hicks1993,Dresselhaus2007}, nanostructuring has been the preferred strategy to enhance the scattering of phonons and hence decrease the lattice thermal conductivity, thus providing an effective strategy to improve the figure of merit of thermoelectric materials. Refinements of this approach, in which defects of different length scales are introduced in order to scatter phonons with a wide range of mean free paths, have led to record high values of $ZT$ up to 2 \cite{Biswas2012,Zhao2014b}. Interestingly, this was achieved by controlling vacancy concentration and grain size. These two are defects that, under control or not, are always present in the sample. In the following sections we report on simulations carried out with the present model to determine the effect of defects on the lattice thermal conductivity.
Notice that from here on we will focus exclusively on the lattice contribution to the thermal conductivity and therefore we will drop the subscript and denote it simply as $\kappa$, which should not to be confused with the total lattice thermal conductivity, including also the electronic contribution.

% Thermal properties of PbTe are extremely interesting because of its low thermal conductivity that makes this material suitable for thermal management applications. Although this material is one of the leading thermoelectric material at intermediate temperatures, it is possible to improve its performance by reducing its lattice thermal conductivity. Including defects into the crystal is one of the most feasible ways to achieve a low lattice thermal conductivity, as well as techniques such as nanostructuring or modeling near a phase transition \cite{savicmurphy1,savicmurphy2}. Here, we will study the effect of defects and grain boundaries on the lattice thermal conductivity of PbTe.

\subsection{Point defects}\label{sec:vacancies} 

Here we examine the effect of point defects on the thermal conductivity in bulk crystalline PbTe using MD simulations and the potential introduced in section \ref{sec:methods}. Intrinsic point defects in a crystal can be vacancies, interstitials or a combination of these ({\it e.g.} Frenkel defects). In PbTe, a relatively small deviation from stoichiometry would result in large concentrations of free charge carriers, something that would be detrimental for the thermoelectric efficiency, and is known to produce aggregates of the excess element in a nearly stoichiometric matrix \cite{Schenk1988}. Therefore stoichiometric defects are probably the most abundant in this material \cite{Schenk1988,Li2015}, and we decided to focus our attention on these. According to this potential, the vacancy formation energies of Schottky defects are in good agreement with first-principles calculations and improve considerably the values obtained in \cite{Qiu2012}: $2.53$ and $4.07$ eV for Schottky dimers and pairs respectively (see table~\ref{table:fittingparams}).

\begin{figure}[h!]
\centering
\includegraphics[width=\columnwidth]{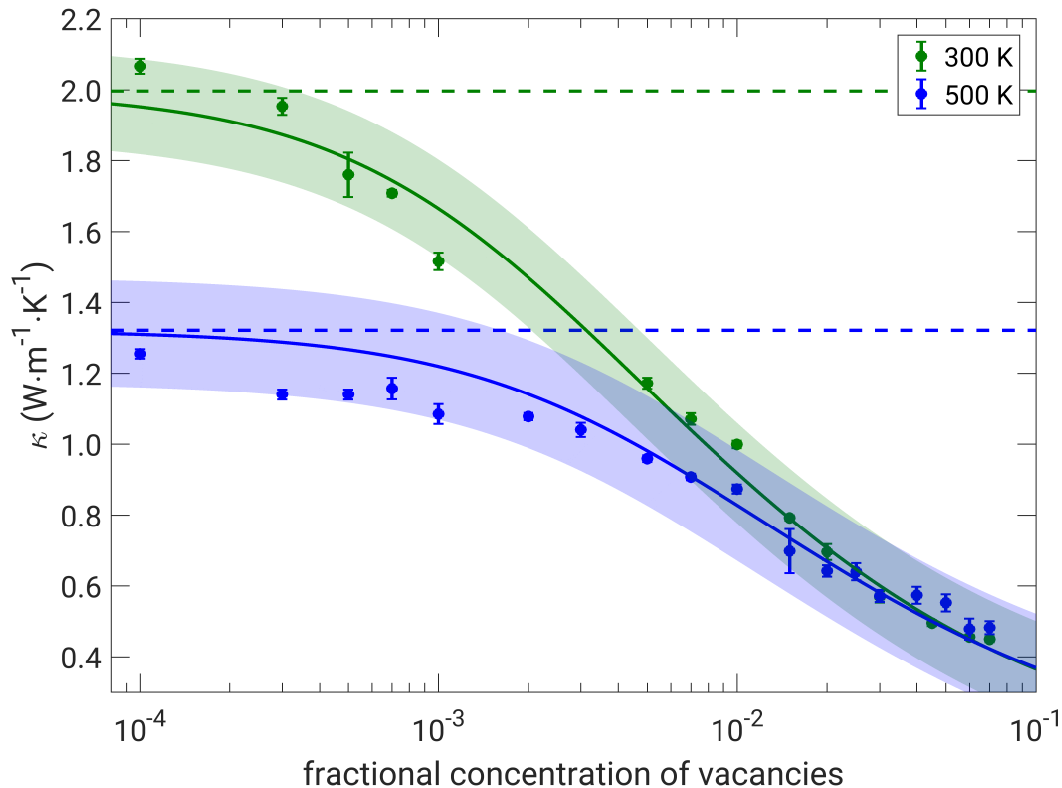}
\caption{Lattice thermal conductivity as a function of vacancy concentrations at 300K and 500 K. Continuous lines represent a fit to the phenomenological model described in \cite{Klemens1960}. Dashed lines are the limit of zero vacancy concentration. Error bars and shadowed regions show the $2\sigma$ confidence interval of the data points and fit respectively.} 
\label{fig:vacanciesrandomboth}
\end{figure}

We first analyze the effect of vacancies by removing atoms from random positions while preserving the stoichiometry of the crystal. The dependence of the lattice conductivity with the concentration of these defects can be seen in figure \ref{fig:vacanciesrandomboth}. 
We observe that at relatively high temperatures (500 K)  the conductivity remains approximately constant over a wide range of vacancy concentrations and it only starts decreasing for concentrations larger than $10^{-3}$. When this happens though, the decrease in conductivity is very pronounced, dropping by $\sim50\%$ for a concentration of $\sim 10^{-2}$.
Conversely, at lower temperatures such as room $T$, we find that the effect of vacancies kicks in at much lower concentrations. 
These results can be compared with the predictions of the analytical model for a Debye solid \cite{Klemens1960}. 
According to the model described in \cite{Yang2004} the conductivity of a crystal in presence of point defects should follow the law
\begin{eqnarray}
{\kappa \over \kappa_0} =  {\arctan\left( u \right) \over u},\quad 
u^2 = {\pi^2 \theta_\mathrm{D} \Omega \over h v^2} \kappa_0 \Gamma
\label{eq:modelDefects}
\end{eqnarray}
where $\kappa_0$ is the conductivity of the pristine crystal, $\theta_\mathrm{D}$ is the Debye temperature, $\Omega$ is the average volume per atom and $v$ is the average sound velocity. The effect of defect concentration enters in (\ref{eq:modelDefects}) through the disorder scaling parameter $\Gamma$, which in presence of point defects may have two contributions, $\Gamma_M$ and $\Gamma_S$, due to mass and strain disorder respectively
\cite{Klemens1955}. For the case of vacancies, the mass disorder rescaling parameter takes the form \cite{Yang2004}:
\begin{eqnarray}
    \Gamma_M = {1\over 2} \sum_{i=\mathrm{Pb, Te}} \left(\overline{M^i} \over \overline{M} \right)^2
               c^i(1-c^i) \left(M^i \over \overline{M^i} \right)^2,
    \label{eq:Gamma_M}
\end{eqnarray}
where the superscript $i$ sums over the two superlattices of Pb and Te atoms, 
$M^i$ is the atomic mass of species $i$, $c^i$ is the concentration of vacancies in the corresponding sublattice, $\overline{M^i}=(1-c^i)M^i$ is the mean atomic mass of sublattice $i$ and $\overline{M}=\overline{M^\mathrm{Pb}}+\overline{M^\mathrm{Te}}$ is the atomic mass of the material averaged over all atomic sites. Similarly, the strain disorder parameter can be expressed as 
\begin{eqnarray}
    \Gamma_S = {1\over 2} \sum_{i=\mathrm{Pb, Te}} \left(\overline{M^i} \over \overline{M} \right)^2
               c^i(1-c^i) \varepsilon^i \left(r^i \over \overline{r^i} \right)^2,
    \label{eq:Gamma_S}
\end{eqnarray}
with $\overline{r^i}=(1-c^i)r^i$ and where $r^i$ are the ionic radii.  (\ref{eq:Gamma_S}) contains an extra rescaling factor, $\varepsilon$, that is a measure of the anharmonicity of the bonds around the point defect \cite{Abeles1963}. Since for vacancies in the rock-salt structure all broken bonds are the same we assume a single value of $\varepsilon$ for both sublattices.

Expressions (\ref{eq:modelDefects})--(\ref{eq:Gamma_S}) have been used to fit the data in figure \ref{fig:vacanciesrandomboth} using only $\varepsilon$ as a fitting parameter; the other parameters have been obtained using the classical potential presented here and are reported in table \ref{table:fittingparams} and the supplementary material.
The fitted curves in figure \ref{fig:vacanciesrandomboth} (solid lines) show that the observed behavior is fully accounted for by the model and that the convergence of curves for different temperatures is the result of scattering by defects becoming the dominant mechanism for thermal resistivity at high defect concentration \cite{Klemens1960}. The values of $\varepsilon$ that we obtain from the fittings, $35\pm3$ and $20\pm2$ for 300 K and 500 K respectively, are similar to values obtained experimentally for traditional semiconductor alloys \cite{Abeles1963}, indicating that the presence of vacancies do not significantly contribute to the anharmonicity of the material.

\begin{figure}[h!]
\centering
\includegraphics[width=\columnwidth]{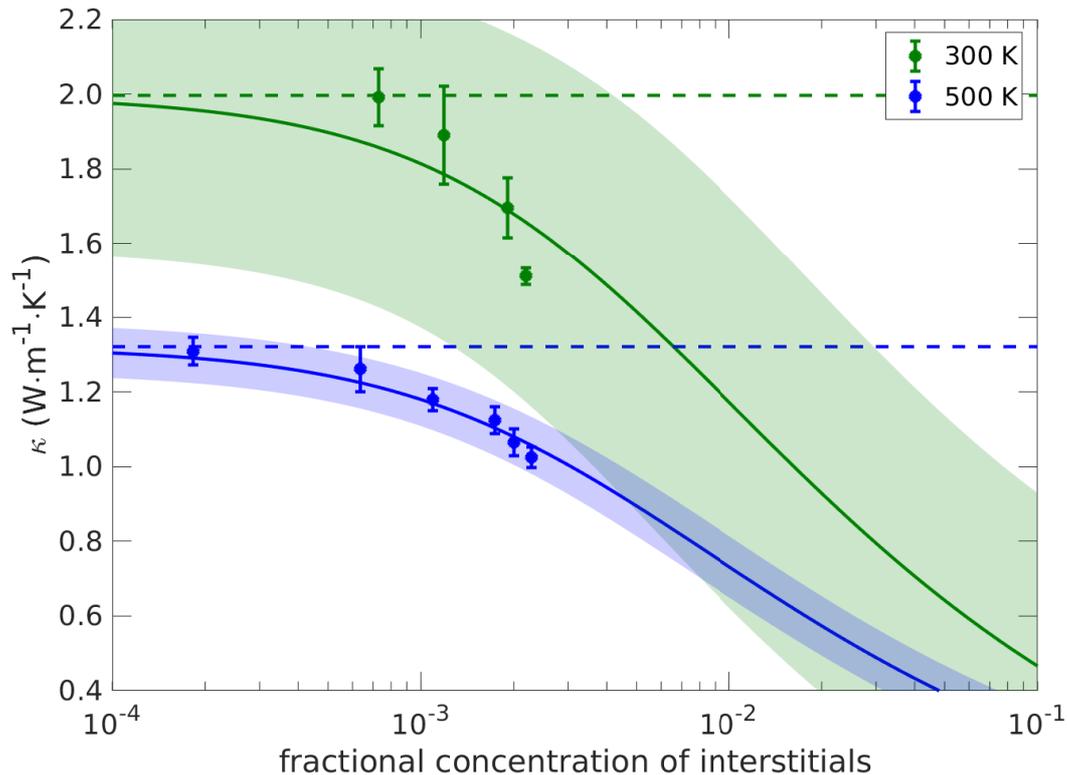}
\caption{Lattice thermal conductivity as a function of the number of interstitial defects at different temperatures. Continuous lines represent a fit to the phenomenological model described in \cite{Klemens1960}. Dashed lines are the limit of zero vacancy concentration. Error bars and shadowed regions show the $2\sigma$ confidence interval of the data poins and fit respectively.} 
\label{fig:addedrandomboth}
\end{figure}

We also explored the effect of interstitial defects, shown in figure \ref{fig:addedrandomboth}. In this case we also maintain the stoichiometry of the material, adding an equal number of Pb and Te atoms.
For the same concentration of interstitial defects as in the case of vacancies, the model described by  (\ref{eq:modelDefects}) predicts a similar decrease in thermal conductivity. This is expected since in both cases, for $c <<1$, we have $\epsilon \sim c$. However, we find that in our MD simulations we cannot reach concentrations of interstitial defects large enough to see the dramatic drop in conductivity observed in the case of vacancies. %since the structure of the material becomes unstable.
%\pa{what do we mean by unstable}
%\jk{Jorge to take care of this}
In reality, large concentrations of interstitials are not expected anyway, since their formation energy is significantly larger than that of other kinds of defects \cite{Li2015}. 

 We used again  (\ref{eq:modelDefects}) to fit the thermal conductivity in presence of interstitials, in this case considering only the strain disorder. The fittings plotted in figure \ref{fig:addedrandomboth} as solid lines show that the model is also able to reproduce the trend in the lattice conductivity obtained from molecular dynamics simulations, although in this case the smaller number of data points results in a large uncertainty of the fitting parameter thus preventing any quantitative conclusion.  

\subsection{Grain boundaries}\label{sec:grains}

Another common type of defect always present in a real sample are grain boundaries. These planar defects act by scattering phonons of much longer mean free paths, thus complementing the effect of point defects at the shorter scales. In fact, controlling the size of the grains is one of the strategies exploited in the multiscale approach to decreasing the lattice thermal conductivity in PbTe and related materials \cite{Biswas2012,Zhao2014b}. Despite reaching thermal conductivities with $ZT\approx 2$, nanostructuring also has its downfalls. Under operation, ageing leads to grain coarsening with the consequent decrease in conductivity \cite{grossfeld2017}. It is therefore useful to determine the region of grain sizes for which the bulk thermal conductivity is recovered.

\begin{figure}
\centering
\includegraphics[width=0.6\columnwidth, clip=true,width=\columnwidth]{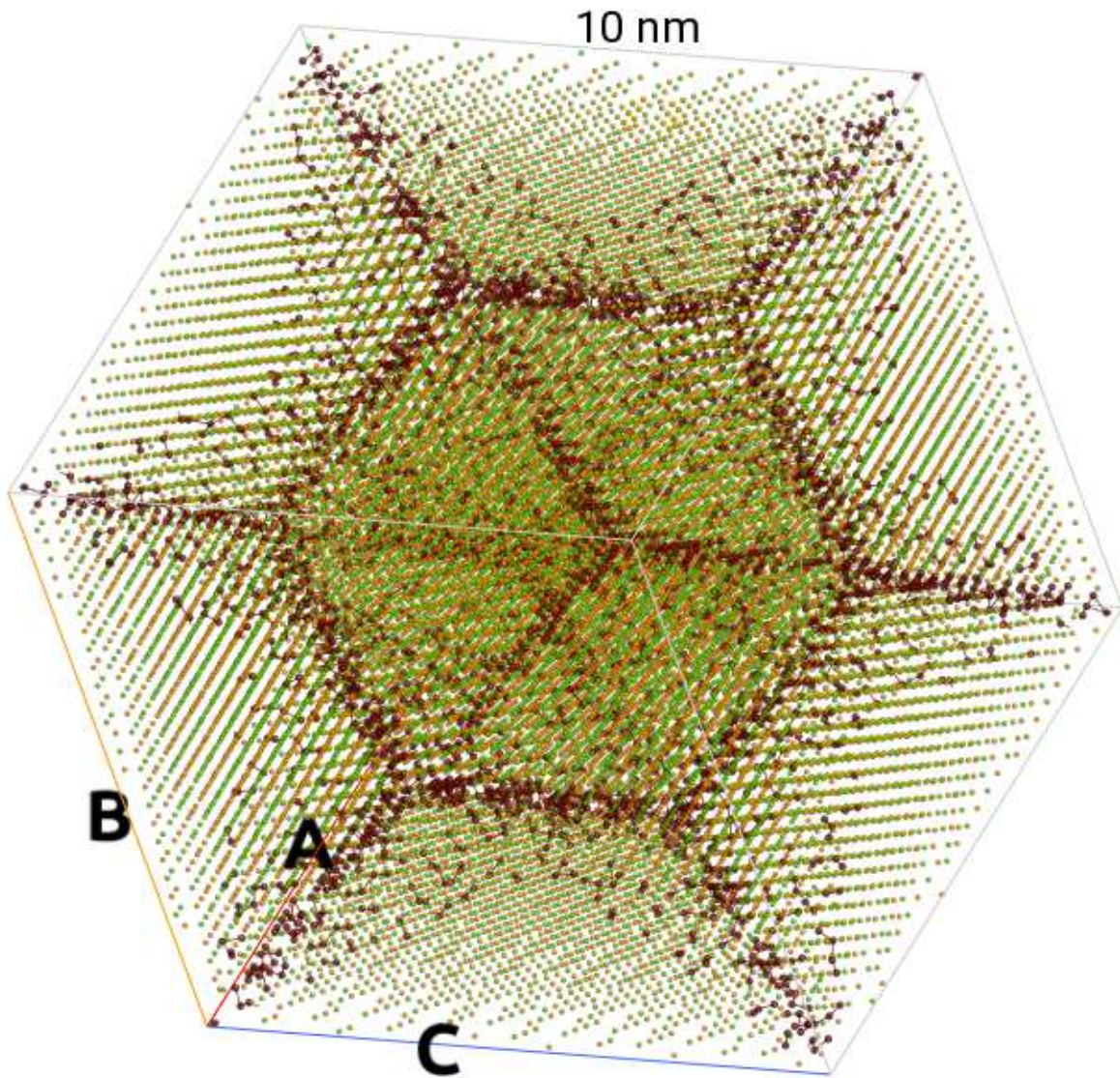}
\caption{Snapshot of a simulation cell with 4 grains. Pb and Te atoms are shown as green and orange balls, respectively. The grain boundaries correspond to the regions in which the distances among atoms cannot be associated with any of the grains and are shown in dark red. Visualization generated with the ATK-ForceField toolkit \cite{ATK-ForceField}.}
\label{fig:grainBox}
\end{figure}
%\pa{Get the number of grains of the cell in figure \ref{fig:grainBox}}
%\pa{What does the color in figure \ref{fig:grainBox} exactly mean?}
%\jf{Javi to take care of this}

In this section, we report on simulations of polycrystalline PbTe carried out in order to assess the effectiveness of phonon scattering by grain boundaries. 
Polycrystalline simulation cells are generated by placing $n$ equally spaced grain seeds on the surface of the simulation box using the ATK-ForceField toolkit \cite{ATK-ForceField}. Then fcc lattices with the orientations of the seeds are grown towards the center of the box, leading to $n$ grains of homogeneous size and allowing for a compact packing of these grains. Total charge is conserved. An example of the resulting structure can be seen in figure \ref{fig:grainBox}. We have studied samples with a number of atoms ranging from 18\,182 to 1\,016\,167, corresponding to grain sizes between 2.68 and 20.47 nm. For the largest samples we have considered $n=4$, while for the smaller ones we used simulation cells with a number of grains between 4 and 32. We also conducted simulations with the same number of grains but different grain size, by varying the volume of the sample.

\begin{figure}[h!]
\centering
\includegraphics[width=\columnwidth]{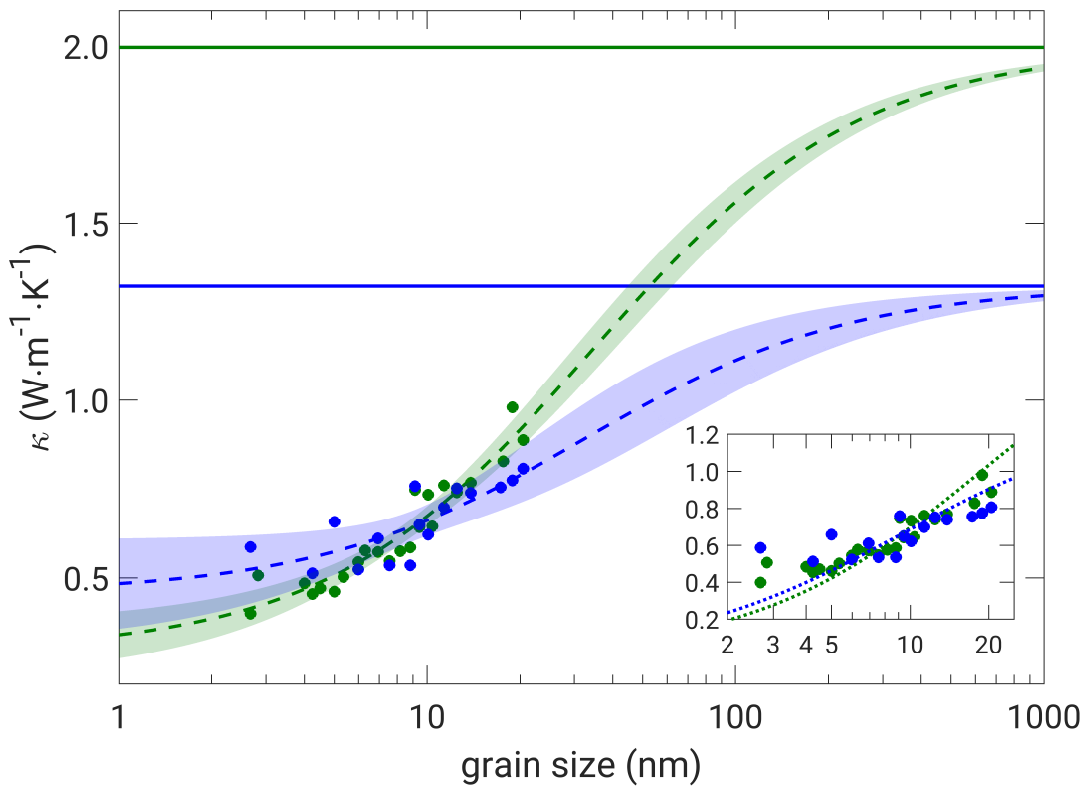}
\caption{Lattice thermal conductivity as a function of particle size in a polycrystaline simulation cell. Results at room temperature (green) and 500 K (blue) are shown. Values obtained using the Green-Kubo formula in MD simulation are plotted as dots, continuous lines correspond to the bulk values of the conductivity and dashed lines represent a fit to the serial model of  (\ref{eq:serial}), where shadow regions showing the $2\sigma$ confidence interval of the fit. The inset shows the same data fitted to a model similar to the one of  (\ref{eq:serial}), but neglecting the finite width of the boundaries. Inset colors as in the main figure.} 
\label{fig:grain}
\end{figure}

The conductivity of the polycrystalline simulation cells as a function of the grain size is displayed in figure \ref{fig:grain} for room temperature and 500 K. The plot shows how the conductivity of the polycrystaline material is drastically reduced with respect to the bulk value in the range of grain sizes considered in our simulations (up to a million atoms). 

These data have been fitted to a simple serial model \cite{Nan1998}, which considers the thermal transport across a series of grains of average size $d$ and internal bulk conductivity $k_b$, separated by boundaries characterized by a Kapitza-type interface resistance $R_\mathrm{K}$,
\begin{eqnarray}
\kappa_\mathrm{eff} = {\kappa_\mathrm{b} \over 
                      1 + R_\mathrm{K}\kappa_\mathrm{b}/d}.
\label{eq:serial1}
\end{eqnarray}
The reduction of the thermal conductivity in polycrystals can be attributed not only to the presence of grain boundaries which hinder the heat transport among grains, but also to the phonon confinement inside individual grains. All phonons with mean free paths greater than the dimensions of the grain will scatter at the boundary and will have a limited contribution to heat transport. 
The effect of this confinement can be accounted for by considering an effective mean free path of $\Lambda_\mathrm{eff}^{-1} = \Lambda_\mathrm{b}^{-1} + (\alpha d)^{-1}$, where $\Lambda_\mathrm{b}$ is the bulk effective mean free path and $\alpha$ is a measure of the transparency of the boundary. Despite being sometimes referred to as transmission coefficient \cite{Wang2011a} it should be noted that values of this parameter are not necessarily bounded between 0 and 1, instead a perfectly transparent boundary is given by $\alpha=\infty$, while $\alpha=0$ yields the completely opaque limit. The assumption $\alpha=1$ is often used, corresponding to the case in which all phonons with mean free path larger than the grain size are truncated \cite{Parrott1969,Wang2011a}. Under this assumption of phonon confinement one obtains the following expression for the effective conductivity of the polycrystaline material \cite{Wang2011a},
%This is formally equivalent to the model described in \onlinecite{Wang2011a},
%
\begin{eqnarray}
\kappa_\mathrm{eff} = {\kappa_\mathrm{b} \over 
                      1 + \Lambda_\mathrm{b}/( \alpha d)}.
\label{eq:serial2}
\end{eqnarray}
It is important to note that both  (\ref{eq:serial1}) and (\ref{eq:serial2}) are formally equivalent and capture in practice the same effects, simply giving a different interpretation to the same phenomenon. The relation between the fundamental parameters of each model can be obtained as 
\begin{eqnarray}
    \alpha = R_\mathrm{K}\kappa_\mathrm{b}/\Lambda_\mathrm{b}.
    \label{eq:serialConn}
\end{eqnarray}
Here we extend the expression in  (\ref{eq:serial1}) to incorporate the finite width of the grain boundary, $\delta$, which had been neglected in previous works, in order to recover the correct limit as the size of the grains tend to zero:
\begin{eqnarray}
\kappa_\mathrm{eff} = {\kappa_\mathrm{b} \over 
                      d/(d+\delta) +
                      R_\mathrm{K}\kappa_\mathrm{b}/(d+\delta)}.
\label{eq:serial}
\end{eqnarray}
%
%These data have been fitted to a simple serial model. The original model, proposed in \cite{Nan1998}, can be extended in \cite{Dong2014} in order to include the effect of the truncation of the mean free path of phonons due to the finite size of the grains. 
%In addition to this, here we incorporate into the model the finite width of the grain boundary, which had been neglected in previous works, in order to recover the correct limit as the size of the grains tend to zero:
%
%\begin{eqnarray}
%\kappa_\mathrm{eff} = {\kappa_\mathrm{b}/(1+\Lambda_\mathrm{b}/d^{0.75}) \over 
%                      d/(d+\delta) +
%                      R_\mathrm{K}\left[\kappa_\mathrm{b}/(1+\Lambda_\mathrm{b}/d^{0.75})\right]/(d+\delta)}.
%\label{eq:serial}
%\end{eqnarray}
%
%In the formula above, $\kappa_\mathrm{eff}$ is the effective conductivity of the polycrystalline material, $\kappa_\mathrm{b}$ is the bulk conductivity, $\Lambda_\mathrm{b}$ is the bulk effective phonon mean free path, $d$ is the average size of the grains, $\delta$ is the effective width of the boundary and $R_\mathrm{K}$ is the Kapitza resistance of the interface. 
%For the fitting we used the values $\Lambda_\mathrm{b}= 7.38$ nm and 5.10 nm for 300 K and 500 K, respectively, obtained from bulk calculations using the direct method \cite{Schelling2002} (see below). 
%
It is worth noting that since we are sampling precisely the range of sizes in which boundaries become the dominant scattering mechanism, including the finite size of the interfaces proved necessary to obtain a reasonable fit. By neglecting this contribution one misses the plateauing of the conductivity at very small sizes, as shown in the inset of figure \ref{fig:grain}.

From the fits shown in figure \ref{fig:grain} we obtain values of $(1.7\pm 0.2)\cdot 10^{-8}$ m$^2$K/W and $(2.4\pm 0.7)\cdot 10^{-8}$ m$^2$K/W for the interface resistance at 300 K and 500 K, respectively. Despite the large uncertainty of these values, the fit suggests that the interface resistance does not depend very strongly on temperature, a result that is supported by the simulations of single grain boundaries described below. 
%
%Comparing  \ref{eq:serial1} and \ref{eq:serial2}, one can obtain the relation $\alpha=R_\mathrm{K}\kappa_\mathrm{b}/\Lambda_\mathrm{b}$. 
Using  (\ref{eq:serialConn}), the values of $R_\mathrm{K}$ obtained from the fittings in figure \ref{fig:grain} and the values of $\Lambda_\mathrm{b}= 7.38$ nm and 5.10 nm for 300 K and 500 K, respectively, obtained from bulk calculations using the direct method \cite{Schelling2002} (see below), we obtain values of $\alpha\sim 0.20$ and 0.16 for 300 K and 500 K, respectively. These values are notably lower than those obtained for nanocrystaline silicon \cite{Wang2011a}, and deviates very significantly from the common assumption of $\alpha=1$.

Due to the relatively large spread of points in figure \ref{fig:grain} the error for the effective width of the boundaries is quite large, with values of $5\pm 1$ nm and $11\pm 5$ nm for 300 K and 500 K respectively. 
%\pa{Maybe a figure to compare by eye with the actual width of a boundary?}
%\jk{Wait and see whether a referee wants it}
%
% Despite the relatively large error bars some qualitative trends can be extracted. The growth with temperature of both the interface scattering and effective width of the boundary can be understood from the increase of square mean displacement with $T$.
%
From the fits to the model one can extrapolate that the conductivity of the polycrystalline material starts drifting away from the bulk value for grains with $d\lesssim 1~\mu$m at 300 K and $d\lesssim 200$ nm at 500 K. On the other end, for small grain sizes, with $d\lesssim 10$ nm, our simulations suggest that the conductivity saturates at a value of $\sim 0.5$ W m$^{-1}$ K$^{-1}$, independently of temperature.  

\begin{figure}[h!]
\centering
\includegraphics[width=\columnwidth]{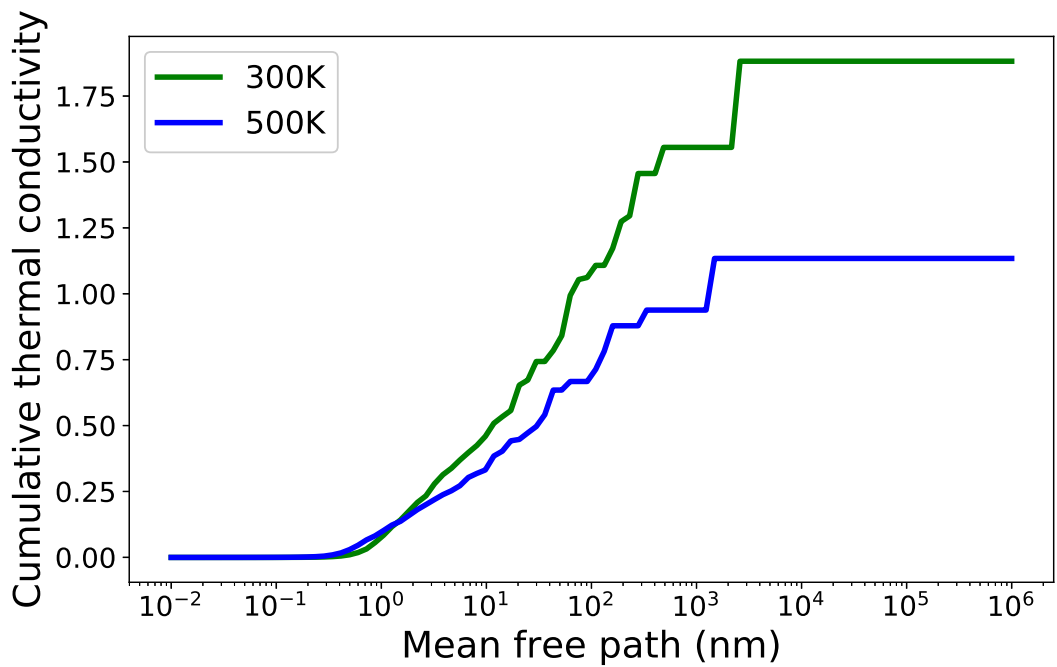}
\caption{ Accumulated thermal conductivity (in W/m K) with respect to phonon mean free path in bulk PbTe at 300 and 500 K.} 
\label{fig:accumulated}
\end{figure}

%{\color{red} This reduction of the thermal conductivity in polycrystals can be attributed not only to the presence of grain boundaries which hinder the heat transport among grains, but also to the phonon confinement inside individual grains. This is why all phonons with mean free paths greater than the dimensions of the grain will scatter at the boundary and will have a limited contribution to heat transport. The phonon Boltzmann transport equation was solved to calculate the lattice thermal conductivity and validate our calculations. 
The validity of phonon confinement interpretation can be tested by comparing our results on the polycrystal with the contribution of phonons to thermal conductivity as a function of their mean free path. This has been calculated solving the Boltzmann transport equation using the ShengBTE code \cite{ShengBTE_2014} with a $6\times6\times6$ supercell and a $20\times20\times20$  grid. The method predicts a bulk  thermal conductivity of $1.88$ W/m K at 300 K and $1.13$ W/m K at 500 K, which are 6\% and 14\% smaller than the value predicted with the Green-Kubo method but within the margin of error usually found in the comparison between both methods. The accumulated thermal conductivity with respect to phonon mean free path in PbTe bulk at 300 and 500 K is presented in figure \ref{fig:accumulated}. % It is observed that phonons with mean free paths smaller than $10$ nm comprise around $100 \%$ of the lattice thermal conductivity for PbTe, what is not in agreement with the results shown in figure \ref{fig:grain}. This means that the reduction of the thermal conductivity in polycrystalline materials is due to the interface resistance rather than the phonon confinement.%
It is observed that not more than $\approx$ 25\% of the thermal conductivity of PbTe is contributed by phonon modes with MFP less than 10 nm at 300K, consistently with the results shown in figure \ref{fig:grain}, in which $\kappa$ is seen to approach the saturation value of 0.5 W/m K for decreasing size.

In order to gain additional insight into the thermal transport properties of the grain boundary we also performed simulations of a single interface between two grains. In this case we used the direct method \cite{Schelling2002} to calculate the lattice thermal conductivity along the direction perpendicular to the boundary between two fcc regions. In the direct method, the thermal conductivity is computed by determining the steady-state temperature gradient at a fixed external heat current between a heat source and heat sink of finite width, defined within the simulation cell. Using an MD time step of $1$ fs, the system was first equilibrated in the NVT ensemble for 200 ps and then under NVE conditions for $200$ ps. At this point the heat source and sink were switched on and the MD simulation continued in the NVE ensemble for $1-5$ ns, depending on the cell size, until reaching a stable temperature profile. A typical temperature profile in the presence of a grain boundary is shown in figure S5 of the supplementary material. The process was repeated for different box lengths ranging up to $130$ nm.
%\jf{Simulations details have been included: timestep=1fs, 2x200m+1.2M timesteps.}

%
The scattering processes at the heat source and sink are expected to contribute very significantly to the thermal conductivity. This effect could be ignored if the simulation cell is longer that the phonon mean free path, but this is often not the case.
Fortunately, the fact that the finite-size simulation cell acts independently from other scattering mechanisms allows us to consider the dependency of the conductivity on the cell length, $L$, according to \cite{Schelling2002}:
\begin{eqnarray}
\frac{1}{\kappa(L)}=\frac{1}{\kappa_\mathrm{b}}\left(1+\frac{4 \Lambda_\mathrm{b}}{L}\right),
\label{eq:direct1}
\end{eqnarray}
where $\kappa_\mathrm{b}$ is the lattice thermal conductivity in bulk and $\Lambda_\mathrm{b}$ is the mean free path in bulk. Performing simulations for various cell lengths and using  (\ref{eq:direct1}) allows us to estimate the bulk conductivity and mean free path, with values $\kappa_\mathrm{b}=2.03$ W m$^{-1}$ K$^{-1}$ and $\Lambda_\mathrm{b}=7.38$ nm at $300$ K and  $\kappa_\mathrm{b}=1.43$ W m$^{-1}$ K$^{-1}$ and $\Lambda_\mathrm{b}=5.10$ nm at $500$ K. In the presence of a grain boundary, a temperature jump at the interface is present and this formula needs to be modified as 
\begin{eqnarray}
\frac{1}{\kappa(L)}=\frac{1}{\kappa_\mathrm{b}}\left(1+\frac{4 \Lambda_\mathrm{b}}{L} + {R_\mathrm{K} \kappa_\mathrm{b} \over L}\right).
\label{eq:direct2}
\end{eqnarray}
Using the values of $\kappa_\mathrm{b}$ and $\Lambda_\mathrm{b}$ calculated in the bulk simulation with the direct method, the magnitude of the interface thermal resistance $R_\mathrm{K}$ can be extracted from calculations at various lengths of the simulation cell. The finite-size scaling is shown in figure S6 of the supplementary material. In addition to  (\ref{eq:direct2}), the interface thermal resistance can be computed from the temperature drop at the boundary, $\Delta T$ as \cite{Crocombette2009}
%\textcolor{red}{(can we give values of these two parameters?)}%\textcolor{red}{(we need to clarify this)}

%
\begin{eqnarray}
R_\mathrm{K}  = {\Delta T \over J},
\label{eq:direct3}
\end{eqnarray}
where $J$ is the constant heat flux imposed between the heat source and sink. We have confirmed that both methods yield the same value of $R_\mathrm{K}$ (within error bars). 

\begin{figure}[h!]
\centering
\includegraphics[width=\columnwidth]{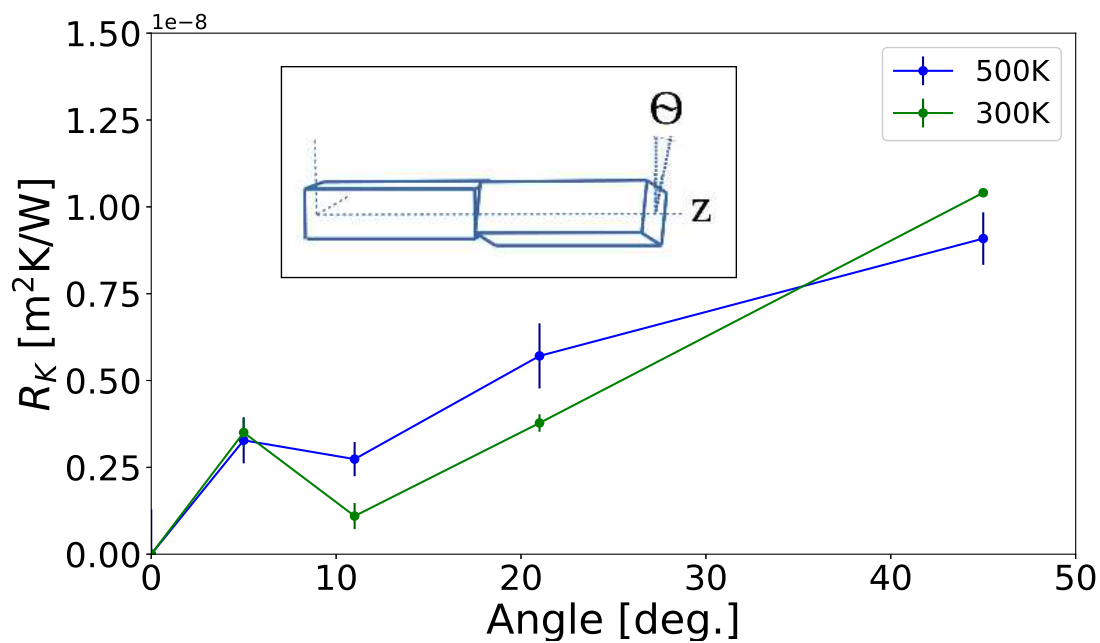}
\caption{Thermal resistance of the a grain boundary as a function of the relative angle of rotation between the crystal structure of two adjacent grains (twist boundary). Inset shows the relative orientation of the two crystal structures.} 
\label{fig:interface}
\end{figure}

\begin{figure}[h!]
\centering
\includegraphics[width=\columnwidth]{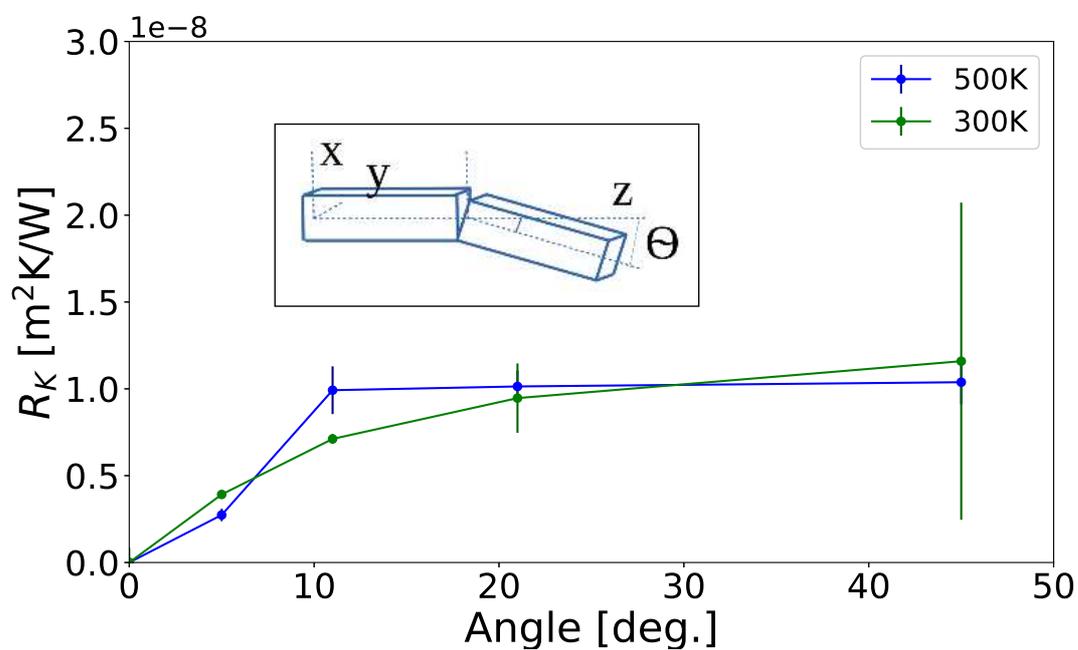} 
\caption{Thermal resistance of the a grain boundary as a function of the relative tilting between the crystal structure of two adjacent grains (tilt boundary). Inset shows the relative orientation of the two crystal structures.} 
\label{fig:interface2}
\end{figure}

We have performed calculations to obtain the grain boundary resistance as a function of the tilting between the two bulk regions. One of the regions is terminated on a [100] surface, which according to density functional theory calculations is the most stable surface in PbTe \cite{Deringer2013}. The crystal structure in the other grain is rotated with respect to the reference one around an axis either perpendicular to the interface (Figure \ref{fig:interface}) or parallel to it (Figure \ref{fig:interface2}). The interface between the two regions is allowed to equilibrate before the non-equilibrium simulation is performed. 

In figure \ref{fig:interface} and \ref{fig:interface2} we plot the values of the thermal resistance of the boundary as a function of the angle for the two types of relative orientations. We observe that for a rotation around an axis perpendicular to the interface, $R_\mathrm{K}$ grows almost linearly with the angle to a maximum value of $\sim 10^{-8}$ m$^2$K/W at $45^{\circ}$. Instead, in the case of a tilting around an axis contained in the boundary plane, the interface resistance grows much more rapidly with the mismatch angle, and reaches the maximum value of, again, $\sim 10^{-8}$ m$^2$K/W for angles $\gtrsim 20^\circ$. 
In both cases we see a negligible difference between results at 300 and 500 K, in line with the results of the polycrystalline simulation. Moreover, a significant decrease in the resistance is observed in presence of small angles, what explains the rise of the thermal conductivity in bulk volume and the lower values in presence of boundaries (Figure \ref{fig:grain}). Despite the fact that these grain boundaries are not necessarily the same as those in the polycrystalline samples, and that the method for determining the thermal resistance is different, the values obtained for $R_\mathrm{K}$ in these two situations are quite close.

\section{Discussion}\label{sec:dicussion} 

The concentration of intrinsic point defects can be estimated from electrical transport methods \cite{Schenk1988} and optical measurements \cite{Zhang2014a}. Reported values in experimental samples typically range from $10^{-5}$ to $10^{-3}$. It should be noted that $10^{-3}$ is also roughly the limit of a deviation from stoichiometry that can be reached before precipitates of the excess element begin to form \cite{Brebrick1960,Schenk1988,Bajaj2015}. In view of our results in section \ref{sec:vacancies}, this suggests several observations. On the one hand, control over growth conditions should permit accessing the onset of the drop of conductivity playing with intrinsic defects alone. However, the benefits of this are limited to samples quenched at relatively low temperature applications, since at high temperatures this kind of defects would attain much lower equilibrium concentrations \cite{Bajaj2015,Li2015}. 
On the other hand, dopants can be introduced in concentrations of up to few percent without disrupting the structure of PbTe \cite{Biswas2012,Pei2011b}, and the reduction in lattice thermal conductivity obtained in figure \ref{fig:vacanciesrandomboth} agrees well with the results reported for extrinsic dopants \cite{Biswas2012,Pei2011b,Zhao2014b}. 
%\pa{we cite too many reviews, maybe cite the sources}
%\jf{Javi to check the reviews for original references}

Our simulations of polycrystalline PbTe show that extremely low conductivity values can be reached for grain sizes in the range of the nanometers. Experimentally though, these grain sizes are very difficult to reach. In fact, it has been determined that down to a fraction of a micrometer the lattice thermal conductivity remains unaffected by grain size \cite{Yoneda2001,Kishimoto2002}, and that the effect of grain size starts to become noticeable for sizes of a few tens of nanometers \cite{Kuo2011,Yoon2013}. In macroscopic samples, smaller grains may not be able to survive the working temperatures of a thermoelectric material. However the regime we explore in section \ref{sec:grains} can be accessed in polycrystalline thin films. Indeed, for Bi$_2$Te$_3$ thin films with grains down to 1 nm in size a drop in conductivity of about 80\% has been measured \cite{Takashiri2008}.
%\pa{maybe mention also the case of precipitates? and also effect of density/voids}

Importantly, for future applications of the force field proposed here and for the development of future parameterizations, our results show that the intrinsic bulk conductivity of PbTe is accessible in experiments, despite de innevitable presence of intrinsic defects. The role of defects has been invoked sometimes to explain the disagreement between MD simulations and experimental results, but we have demonstrated here that the most common defects only play a role in extreme regimes and that most experimental reports of the bulk PbTe lattice conductivity do really probe the intrinsic bulk value. 

\section{Conclusions}\label{sec:conclusions} 

We have calculated the lattice thermal conductivity of PbTe containing the most common types of intrinsic defects. For this we have presented a new parametrization of a Coulomb-Buckingham potential that reproduces very accurately the evolution of the lattice thermal conductivity over a wide range of temperatures. Using this potential we found that both the presence of vacancies and grain boundaries can separately reduce the value of the thermal conductivity to $\sim 0.5$ W/mK. In the regimes of high density of point or planar defects the conductivity becomes practically independent of temperature, as predicted by phenomenological models. 
Extrinsic defects, such as dopants or precipitates are know to have a similar effect, but they may interfere with the electronic properties of the system. Although this interference can sometimes be used to our advantage, the results of this work provide complementary strategies to optimize the performance of thermoelectric devices. Further improvements can be achieved by introducing extrinsic defects that drive the material close to a ferroelectric phase transition, e.g. by alloying with Ge, but this is outside the scope of the present study, which focuses on PbTe, rather than its alloys\cite{Murphy2016}.
Finally, our study demonstrates that experimental values of the conductivity available in the literature do correspond to the bulk value, since the effect of imperfections only becomes noticeable at  relatively high concentrations of point defects and grain boundaries. This highlights the importance of having a classical potential capable of reproducing accurately the conductivity of the bulk material, something that has to be taken into consideration when developing new parameterizations.

\ack
This work was supported by a research grant from Science Foundation 
Ireland (SFI) and the Department for the Economy Northern Ireland under 
the SFI-DfE Investigators Programme Partnership, Grant Number 15/IA/3160. We are grateful for computational support from the UK national high performance computing service, ARCHER, for which access was obtained via the UKCP consortium and funded by EPSRC grant ref EP/P022561/1, and from the UK Materials and Molecular Modelling Hub, which was partially funded by EPSRC grant ref EP/P020194/1. We thank Tchavdar Todorov, Myrta Gr\"uning, Piotr Chudzinski and G. Jeffrey Snyder for insightful discussions.

% MD simulations have been performed to reproduce the low lattice thermal conductivity of bulk PbTe at intermediate temperatures. This low thermal conductivity make this material one of the most important thermoelectric materials, but new strategies to improve its performance are under research. Due to the relation between most of the magnitudes that define the efficiency of these materials, it is not straightforward to find ways to enhance the efficiency, but one of them may be to reduce the lattice thermal conductivity. In this paper, we studied several strategies which enhance the performance of PbTe using molecular dynamics simulations.

% We saw that simple classical potentials, like a Buckingham potential, allow calculating the thermal conductivity of PbTe in the temperature interval between 300 and 800K. This potential was used to analyze different strategies to reduce the thermal conductivity using both the Green-Kubo method and the direct method, and the best results were obtained when we remove atoms from the simulation cell at random positions or when small grains are created. In both cases, the thermal conductivity falls by up to a $75 \%$, and in the second case it may be due to the existence of a grain boundary sliding. 

\section*{References}
\providecommand{\newblock}{}


\begin{thebibliography}{10}
\expandafter\ifx\csname url\endcsname\relax
  \def\url#1{{\tt #1}}\fi
\expandafter\ifx\csname urlprefix\endcsname\relax\def\urlprefix{URL }\fi
\providecommand{\eprint}[2][]{\url{#2}}

\bibitem{Schierning2018}
Schierning G 2018 {\em Nature Energy\/} {\bf 3} 92--93 ISSN 20587546
  \urlprefix\url{http://dx.doi.org/10.1038/s41560-018-0093-4}

\bibitem{Champier2017}
Champier D 2017 {\em Energy Conversion and Management\/} {\bf 140} 167--181
  ISSN 01968904
  \urlprefix\url{http://dx.doi.org/10.1016/j.enconman.2017.02.070}

\bibitem{Schmidt2009}
Schmidt G~R, Sutliff T~J and Dudzinski L~A 2009 {\em 60th International
  Astronautical Congress 2009, IAC 2009\/} {\bf 8} 6612--6633 ISSN AIAA
  2008-5640
  \urlprefix\url{http://www.scopus.com/inward/record.url?eid=2-s2.0-77953523923{\&}partnerID=tZOtx3y1}

\bibitem{Snyder2008}
Snyder G~J and Toberer E~S 2008 {\em Nature materials\/} {\bf 7} 105--114

\bibitem{Delaire2011}
Delaire O, Ma J, Marty K, May a~F, McGuire M~a, Du M~H, Singh D~J, Podlesnyak
  A, Ehlers G, Lumsden M~D and Sales B~C 2011 {\em Nature Materials\/} {\bf 10}
  614--619 ISSN 1476-1122 \urlprefix\url{http://dx.doi.org/10.1038/nmat3035
  http://www.ncbi.nlm.nih.gov/pubmed/21642983{\%}5Cnhttp://www.nature.com/doifinder/10.1038/nmat3035}

\bibitem{An2008a}
An J, Subedi A and Singh D~J 2008 {\em Solid State Communications\/} {\bf 148}
  417--419 ISSN 00381098 (\textit{Preprint} \eprint{0806.2727})
  \urlprefix\url{http://dx.doi.org/10.1016/j.ssc.2008.09.027}

\bibitem{Shiga2012}
Shiga T, Shiomi J, Ma J, Delaire O, Radzynski T, Lusakowski A, Esfarjani K and
  Chen G 2012 {\em Phys. Rev. B\/} {\bf 85} 155203 ISSN 10980121

\bibitem{Ju2018}
Ju S, Shiga T, Feng L and Shiomi J 2018 {\em Physical Review B\/} {\bf 97}
  184305 ISSN 24699969

\bibitem{Cochran1966}
Cochran W, FRS, Cowley R~A, Dolling G and Elcombe M~M 1966 {\em Proc. R. Soc.
  Lond. A\/} {\bf 293} 433 ISSN 1364-5021

\bibitem{Pei2011b}
Pei Y, LaLonde A, Iwanaga S and Snyder G~J 2011 {\em Energy {\&} Environmental
  Science\/} {\bf 4} 2085 ISSN 1754-5692
  \urlprefix\url{http://xlink.rsc.org/?DOI=c0ee00456a}

\bibitem{Heremans2008}
Heremans J~P, Jovovic V, Toberer E~S, Saramat A, Kurosaki K, Charoenphakdee A,
  Yamanaka S and Snyder G~J 2008 {\em Science\/} {\bf 321} 554--557 ISSN
  0036-8075
  \urlprefix\url{http://www.sciencemag.org/content/321/5888/554.abstract{\%}5Cnhttp://www.sciencemag.org/cgi/doi/10.1126/science.1159725}

\bibitem{Pei2011}
Pei Y, Shi X, LaLonde A, Wang H, Chen L and Snyder G~J 2011 {\em Nature\/} {\bf
  473} 66--69 ISSN 0028-0836
  \urlprefix\url{http://www.nature.com/doifinder/10.1038/nature09996}

\bibitem{Pei2011c}
Pei Y, Lensch-Falk J, Toberer E~S, Medlin D~L and Snyder G~J 2011 {\em Advanced
  Functional Materials\/} {\bf 21} 241--249 ISSN 1616301X

\bibitem{Biswas2012}
Biswas K, He J, Blum I~D, Wu C~I, Hogan T~P, Seidman D~N, Dravid V~P and
  Kanatzidis M~G 2012 {\em Nature\/} {\bf 489} 414--8 ISSN 1476-4687
  \urlprefix\url{http://www.ncbi.nlm.nih.gov/pubmed/22996556}

\bibitem{Zhao2014b}
Zhao L~D, Dravid V~P and Kanatzidis M~G 2014 {\em Energy Environ. Sci.\/} {\bf
  7} 251--268 ISSN 1754-5692
  \urlprefix\url{http://xlink.rsc.org/?DOI=C3EE43099E}

\bibitem{Murphy2016}
Murphy R~M, Murray E~D, Fahy S and Savic I 2016 {\em Physical Review B\/} {\bf
  93} 104304 ISSN 1550235X

\bibitem{Murphy2017}
Murphy R~M, Murray {\'{E}}~D, Fahy S and Savi{\'{c}} I 2017 {\em Physical
  Review B\/} {\bf 95} 144302 ISSN 2469-9950
  \urlprefix\url{http://link.aps.org/doi/10.1103/PhysRevB.95.144302}

\bibitem{Li2015}
Li W~F, Fang C~M, Dijkstra M and {Van Huis} M~A 2015 {\em Journal of Physics
  Condensed Matter\/} {\bf 27} 355801 ISSN 1361648X

\bibitem{Schenk1988}
Schenk M, Berger H, Klimakow A, M{\"{u}}hlberg M and Wienecke M 1988 {\em
  Crystal Research and Technology\/} {\bf 23} 77--84 ISSN 15214079

\bibitem{Bajaj2015}
Bajaj S, Pomrehn G~S, Doak J~W, Gierlotka W, Wu H~J, Chen S~W, Wolverton C,
  Goddard W~A and {Jeffrey Snyder} G 2015 {\em Acta Materialia\/} {\bf 92}
  72--80 ISSN 13596454
  \urlprefix\url{http://dx.doi.org/10.1016/j.actamat.2015.03.034}

\bibitem{Goyal2017}
Goyal A, Gorai P, Toberer E~S and Stevanovi{\'{c}} V 2017 {\em npj
  Computational Materials\/} {\bf 3} 42 ISSN 20573960

\bibitem{Chonan2006}
Chonan T and Katayama S 2006 {\em Journal of the Physical Society of Japan\/}
  {\bf 75} 064601 ISSN 0031-9015
  \urlprefix\url{http://journals.jps.jp/doi/abs/10.1143/JPSJ.75.064601}

\bibitem{Qiu2012}
Qiu B, Bao H, Zhang G, Wu Y and Ruan X 2012 {\em Computational Materials
  Science\/} {\bf 53} 278--285 ISSN 09270256
  \urlprefix\url{http://dx.doi.org/10.1016/j.commatsci.2011.08.016}

\bibitem{Zhou2018}
Zhou Y, Yang J~J~y, Chen L, Hu M, Cheng L and Hu M 2018 {\em Phys. Rev. B\/}
  {\bf 97} 085304

\bibitem{Kim2012a}
Kim H and Kaviany M 2012 {\em Physical Review B\/} {\bf 86} 045213 ISSN
  10980121

\bibitem{Bozin2010}
Bo{\v{z}}in E~S, Malliakas C~D, Souvatzis P, Proffen T, Spaldin N~a, Kanatzidis
  M~G and Billinge S~J~L 2010 {\em Science\/} {\bf 330} 1660--3 ISSN 1095-9203
  \urlprefix\url{http://www.ncbi.nlm.nih.gov/pubmed/21164012
  http://www.ncbi.nlm.nih.gov/pubmed/21164012{\%}5Cnhttp://www.sciencemag.org/content/330/6011/1660}

\bibitem{Shiga2014}
Shiga T, Murakami T, Hori T, Delaire O and Shiomi J 2014 {\em Applied Physics
  Express\/} {\bf 7}

\bibitem{Tian2012}
Tian Z, Garg J, Esfarjani K, Shiga T, Shiomi J and Chen G 2012 {\em Physical
  Review B\/} {\bf 85} 184303

\bibitem{Romero2015}
Romero A~H, Gross E~K~U, Verstraete M~J and Hellman O 2015 {\em Physical Review
  B\/} {\bf 91} 214310

\bibitem{Qiu2008}
Qiu B, Bao H and Ruan X 2008 {Multiscale Simulations of Thermoelectric
  Properties of PbTe} {\em ASME2008 3rd Energy Nanotechnology International
  Conference\/} pp 45--60
  \urlprefix\url{http://proceedings.asmedigitalcollection.asme.org/proceeding.aspx?articleid=1626940}

\bibitem{Gale1997}
Gale J~D 1997 {\em JCS Faraday Trans.\/} {\bf 93} 629

\bibitem{Gale2003}
Gale J~D and Rohl A~L 2003 {\em Mol. Simul.\/} {\bf 29} 291

\bibitem{Zhang2011a}
Zhang Y, Ke X, Kent P, Yang J and Chen C 2011 {\em Physical Review Letters\/}
  {\bf 107} 175503 ISSN 0031-9007
  \urlprefix\url{http://link.aps.org/doi/10.1103/PhysRevLett.107.175503}

\bibitem{Gonze1997}
Gonze X and Lee C 1997 {\em Phy. Rev. B\/} {\bf 55} 10355

\bibitem{Powell2006}
Powell D, Migliorato M~A and Cullis A~G 2006 {\em Physica E: Low-Dimensional
  Systems and Nanostructures\/} {\bf 32} 270--272 ISSN 13869477

\bibitem{Giannozzi2009}
Giannozzi P, Baroni S, Bonini N, Calandra M, Car R, Cavazzoni C, Ceresoli D,
  Chiarotti G~L, Cococcioni M, Dabo I, {Dal Corso} A, de~Gironcoli S, Fabris S,
  Fratesi G, Gebauer R, Gerstmann U, Gougoussis C, Kokalj A, Lazzeri M,
  Martin-Samos L, Marzari N, Mauri F, Mazzarello R, Paolini S, Pasquarello A,
  Paulatto L, Sbraccia C, Scandolo S, Sclauzero G, Seitsonen A~P, Smogunov A,
  Umari P and Wentzcovitch R~M 2009 {\em J. Phys.: Condens. Matter\/} {\bf 21}
  395502 \urlprefix\url{http://www.ncbi.nlm.nih.gov/pubmed/21832390}

\bibitem{Monkhorst1976}
Monkhorst H and Pack J 1976 {\em Physical Review B\/} {\bf 13} 5188--5192
  \urlprefix\url{http://www.if.pwr.wroc.pl/{~}scharoch/Abinitio/MonkhorstPack.pdf}

\bibitem{MadelungHandbook_PbTe}
and editors of the~volumes III/17E-17F-41C C~A {Lead telluride (PbTe) crystal
  structure, lattice parameters, thermal expansion} {\em Non-Tetrahedrally
  Bonded Elements and Binary Compounds I\/} vol 41C ed Madelung O,
  R{\"{o}}ssler U and Schulz M (Berlin, Heidelberg: Springer Berlin Heidelberg)
  ISBN 978-3-540-31360-1
  \urlprefix\url{https://doi.org/10.1007/10681727{\_}711}

\bibitem{Miller1981}
Miller A~J, Saunders G~A and Yogurt{\c{c}}u Y~K 1981 {\em J. Phys. C: Solid
  State Phys\/} {\bf 14} 1569

\bibitem{WunFanLi2015}
Wun-Fan~Li Chang-Ming~Fang M~D and van Huis~Soft M~A 2015 {\em J. Phys.
  Condens. Matter\/} {\bf 27} 355801

\bibitem{Zwanzig1964}
Zwanzig R 1964 {\em Annu. Rev. Phys. Chem.\/} {\bf 16} 67--102 ISSN 0066-426X

\bibitem{Che2000}
Che J, {\c{C}}ağın T, Deng W and Goddard W~A 2000 {\em The Journal of
  Chemical Physics\/} {\bf 113} 6888--6900 ISSN 0021-9606
  \urlprefix\url{http://aip.scitation.org/doi/10.1063/1.1310223}

\bibitem{Plimpton1995}
Plimpton S 1995 {\em J. Comp. Phys.\/} {\bf 117} 1

\bibitem{lammpsWeb}
http://lammps.sandia.gov

\bibitem{Sootsman2009}
Sootsman J~R, He J, Dravid V~P, Li C~P, Uher C and Kanatzidis M~G 2009 {\em
  Journal of Applied Physics\/} {\bf 105} 083718 ISSN 00218979

\bibitem{Slack1979}
Slack G~A 1979 {\em Solid State Physics\/} {\bf 34} 1--71

\bibitem{Morelli2002}
Morelli D~T and Heremans J~P 2002 {\em Applied Physics Letters\/} {\bf 81}
  5126--5128 ISSN 00036951

\bibitem{Zhang2009}
Zhang Y, Ke X, Chen C, Yang J and Kent P~R~C 2009 {\em Physical Review B\/}
  {\bf 80} 024304

\bibitem{Hicks1993}
Hicks L~D and Dresselhaus M~S 1993 {\em Phys. Rev. B\/} {\bf 47} 727--731

\bibitem{Dresselhaus2007}
Dresselhaus M~S, Chen G, Tang M, Yang R, Lee H, Wang D, Ren Z, Fleurial J~p and
  Gogna P 2007 {\em Advanced Materials\/} {\bf 19} 1043--1053

\bibitem{Klemens1960}
Klemens P 1960 {\em Physical Review\/} {\bf 119} 507--509 ISSN 0031-899X
  \urlprefix\url{http://link.aps.org/doi/10.1103/PhysRev.119.507}

\bibitem{Yang2004}
Yang J, Meisner G~P and Chen L 2004 {\em Applied Physics Letters\/} {\bf 85}
  1140--1142 ISSN 0003-6951

\bibitem{Klemens1955}
Klemens P~G 1955 {\em Proceedings of the Physical Society. Section A\/} {\bf
  68} 1113--1128 ISSN 03701298

\bibitem{Abeles1963}
Abeles B 1963 {\em Physical Review\/} {\bf 131} 1906--1911 ISSN 0031899X

\bibitem{grossfeld2017}
Grossfeld T, Sheskin A, Gelbstein Y and Amouyal Y 2017 {\em Crystals\/} {\bf 7}
  281 ISSN 2073-4352 \urlprefix\url{http://www.mdpi.com/2073-4352/7/9/281}

\bibitem{ATK-ForceField}
Schneider J, Hamaekers J, Chill S~T, Smidstrup S, Bulin J, Thesen R, Blom A and
  Stokbro K 2017 {\em Modelling Simul. Mater. Sci. Eng.\/} {\bf 25} 085007

\bibitem{Nan1998}
Nan C~W and Birringer R 1998 {\em Physical Review B\/} {\bf 57} 8264--8268 ISSN
  1550235X

\bibitem{Wang2011a}
Wang Z, Alaniz J~E, Jang W, Garay J~E and Dames C 2011 {\em Nano Letters\/}
  {\bf 11} 2206--2213 ISSN 15306984

\bibitem{Parrott1969}
Parrott J~E 1969 {\em Journal of Physics C: Solid State Physics\/} {\bf 2}
  147--151 ISSN 00223719

\bibitem{Schelling2002}
Schelling P~K, Phillpot S~R and Keblinski P 2002 {\em Physical Review B\/} {\bf
  65} 144306 ISSN 1550235X

\bibitem{ShengBTE_2014}
Li W, Carrete J, Katcho N~A and Mingo N 2014 {\em Comp. Phys. Commun.\/} {\bf
  185} 1747–1758

\bibitem{Crocombette2009}
Crocombette J~P and Gelebart L 2009 {\em Journal of Applied Physics\/} {\bf
  106} 083520 ISSN 0021-8979
  \urlprefix\url{http://aip.scitation.org/doi/10.1063/1.3240344}

\bibitem{Deringer2013}
Deringer V~L and Dronskowski R 2013 {\em Journal of Physical Chemistry C\/}
  {\bf 117} 24455--24461 ISSN 19327447

\bibitem{Zhang2014a}
Zhang B, Cai C, Jin S, Ye Z, Wu H and Qi Z 2014 {\em Applied Physics Letters\/}
  {\bf 105} 022109 ISSN 00036951

\bibitem{Brebrick1960}
Brebrick R~F and Allgaier R~S 1960 {\em The Journal of Chemical Physics\/} {\bf
  32} 1826--1831 ISSN 00219606

\bibitem{Yoneda2001}
Yoneda S, Ohta E, Kaibe H~T, Oshugi I~J, Shiota I and Nishida I~A 2001 {\em
  Materials Transactions\/} {\bf 42} 329--335

\bibitem{Kishimoto2002}
Kishimoto K and Koyanagi T 2002 {\em Journal of Applied Physics\/} {\bf 92}
  2544--2549 ISSN 00218979

\bibitem{Kuo2011}
Kuo C~H, Chien H~S, Hwang C~S, Chou Y~W, Jeng M~S and Yoshimura M 2011 {\em
  Materials Transactions\/} {\bf 52} 795--801 ISSN 1345-9678
  \urlprefix\url{https://www.jstage.jst.go.jp/article/matertrans/52/4/52{\_}M2010331/{\_}article}

\bibitem{Yoon2013}
Yoon S, Kwon O~J, Ahn S, Kim J~Y, Koo H, Bae S~H, Cho J~Y, Kim J~S and Park C
  2013 {\em Journal of Electronic Materials\/} {\bf 42} 3390--3396 ISSN
  0361-5235 \urlprefix\url{http://link.springer.com/10.1007/s11664-013-2753-2}

\bibitem{Takashiri2008}
Takashiri M, Miyazaki K, Tanaka S, Kurosaki J, Nagai D and Tsukamoto H 2008
  {\em Journal of Applied Physics\/} {\bf 104} 084302 ISSN 00218979

\end{thebibliography}
\end{document}